\documentclass[11pt]{article}

\usepackage{graphicx}
\usepackage{amsfonts,amsmath,amssymb,bm, amsthm}
\usepackage{natbib}
\usepackage{url}
\usepackage{hyperref}
\hypersetup{colorlinks=false,pdfborder={0 0 0}}

\usepackage[utf8]{inputenc}
\usepackage[english]{babel}

\usepackage[margin = 2.5cm]{geometry}

\usepackage[plain,noend]{algorithm2e}

\def\Pr{{\text{P}}}

\newtheorem{theorem}{Theorem}
\newtheorem{lemma}{Lemma}

\title{Improved small-sample inference for functions of parameters in the $k$-sample multinomial problem}
\author{Michael C Sachs, Erin E Gabriel, Michael P Fay}
\date{\today}

\begin{document}

\maketitle

\begin{abstract}
When the target parameter for inference is a real-valued, continuous function of probabilities in the $k$-sample multinomial problem, variance estimation may be challenging. In small samples or when the function is nondifferentiable at the true parameter, methods like the nonparametric bootstrap or delta method may perform poorly. We develop an exact inference method that applies to this general situation. We prove that our proposed exact p-value correctly bounds the type I error rate and the associated confidence intervals provide at least nominal coverage; however, they are generally difficult to implement. Thus, we propose a Monte Carlo implementation to estimate the exact p-value and confidence intervals that we show to be consistent as the number of iterations grows. Our approach is general in that it applies to any real-valued continuous function of multinomial probabilities from an arbitrary number of samples and with different numbers of categories. \\
\noindent
\textbf{Keywords} --- exact inference; multinomial; computation. %
\end{abstract}%

\section{Introduction}

The target of inference in applied settings is sometimes a 
function of the probability parameters
from one or more multinomial random variables. For example, symbolic nonparametric bounds, such as the well-known Balke-Pearl IV Bounds \citep{Balke97}, are often the maximum or minimum of a set of terms that are the linear combination of probabilities from two or more independent multinomials. Much of the literature on such bounds has focused on their derivation, suggesting maximum likelihood estimators (MLE) and bootstrap standard errors for inference. 

Although it is well-known that the MLE of such a function is the function applied to the empirical proportions, i.e. the MLE of the probabilities of the multinomials, inference can be more difficult. When the function is continuously differentiable, applying the delta method to obtain a variance estimator is straightforward. However, if the function is complicated,
or its variance is otherwise difficult to calculate, then one may consider using the nonparametric bootstrap for inference \citep{efron1979bootstrap}. Unfortunately, either of these variance estimators may perform poorly in small samples or even in large sample sizes when the function is not differentiable at the true parameter or if the parameter is on the boundary of the range \citep{bickel1981some, bickel2008choice}. In the symbolic bounds literature, these problems are often ignored or assumed not to exist. This is, however, counter to the spirit of nonparametric symbolic bounds where the objective is to make the fewest assumptions possible. 

Another example of a parameter that can suffer from the same issues and thus give poor inference is the Bhattacharyya coefficient, which we illustrate in our simulations \citep{sankhya1946measure}. The Bhattacharyya coefficient quantifies the closeness of two multinomial distributions. When the coefficient is one or zero, even in large sample sizes, inference can be improper when based on asymptotic theory that relies on the parameter being bounded away from the boundary. Another example of real-valued parameters are measures of relative preference among a fixed set of categories, such as the difference or ratio between two multinomial parameters within a single sample \citep{nelson1972statistical}. 

Exact methods are of course not new, however, the main focus of previous work in exact methods for multinomial settings has been testing for independence or specific to the function being considered. \citet{agresti1992survey} provides a review of exact methods for contingency tables. \citet{resin2023simple} focuses on fast algorithms to compute exact goodness-of-fit tests. 
\citet{chafai2009confidence} 
and \citet{malloy2021optimal} provide exact confidence regions for a single multinomial parameter vector. 
\citet{frey2009exact} describes an exact equivalence testing method for one sample multinomial settings, which is a problem that fits into our general framework, but their method induces an ordering on the multinomial parameters that is not inherent. Instead, we propose a general method for tests based on exact p-values and confidence intervals, which always provide the correct type I error rate and coverage for any real-valued function of any number of multinomials with any number of levels. Although tests like Fisher's Exact fit into our framework, we do not wish to or claim to improve upon the vast research of the two-sample binomial problem. 

Although we prove that our proposed method provides an exact p-value for testing and correct coverage for confidence intervals regardless of the nature of the problem, including parameter boundary issues or small sample sizes, we also conclude that it is computationally infeasible. We provide a method for computational approximation, which we prove to converge in probability to the exact p-value as the number of iterations of the approximation grows. We describe our implementation and usage of the method in an R package called \texttt{xactonomial}, evaluate it in simulations, and illustrate its use in a real data example using Balke-Pearl causal bounds in the randomized trial setting with noncompliance.

\section{Setting and notation}

We consider the following problem. Let $T_j$ be distributed $\mbox{Multinomial}(n_j, d_j, \theta_j)$ where $n_j$ is the number of trials, $d_j$ the number of mutually exclusive categories, and $\theta_j$ the vector of $d_j$ probabilities for $j = 1, \ldots, k$. Denote $\boldsymbol{T} = (T_1, \ldots, T_k)$ and $\boldsymbol{\theta} = (\theta_1, \ldots, \theta_k)$. Suppose we observe a sample $\boldsymbol{t} = (t_1, \ldots, t_k)$  which is a vector of counts corresponding to the random variable $\boldsymbol{T}$. Let $G(\boldsymbol{t})$ denote a real-valued statistic, where $G(\cdot)$ is used to define an ordering of the sample space.  

Suppose one is interested in the parameter $\psi \equiv \tau(\boldsymbol{\theta}) \in \Psi$ 
where $\Psi$ is a subset of the real line $\mathbb{R}$.  
 An example for $k = 2$ and $d_1 = d_2 = d$ is the Bhattacharyya coefficient between $T_1$ and $T_2$ \citep{sankhya1946measure}. In that case we have $\boldsymbol{\theta} = (\theta_1, \theta_2)$, where $\theta_1, \theta_2$ each have $d$ elements, and $\tau(\boldsymbol{\theta}) = \sum_{i = 1}^d \sqrt{\theta_{1i}\theta_{2i}}$. A simple example for $k = 1$ is $\tau(\boldsymbol{\theta}) = \max_{i \in \{ 1,\ldots, d_1 \} } \{\theta_{1i} \}$, i.e., the largest element of the probability vector. The third example would be  symbolic nonparametric bounds in discrete settings, which often take the form of the $\max$ and $\min$ of a number of terms that are linear combinations of multinomial probabilities \citep{sachs2023general}.  


\section{Exact inference}
In the following we derive the p-values and confidence intervals that we prove are theoretically valid. 

\subsection{P-values}

Consider
testing the null hypothesis, $H_0: \psi \leq \psi_0$ versus the alternative hypothesis,
$H_1: \psi > \psi_0.$ Denote the joint probability mass function $f(\boldsymbol{t}, \boldsymbol{\theta}) = \Pr\{\boldsymbol{T} = \boldsymbol{t} | \boldsymbol{\theta} \}$. 
The parameter space under the null is $\Theta_0(\psi_0) = \{\boldsymbol{\theta}: \tau(\boldsymbol{\theta}) \leq \psi_0\}.$  Let $\mathcal{S}$ denote the sample space, i.e., the set of possible outcomes of the $\boldsymbol{T}$ random vector. Let $\overline{\mathcal{S}} (\boldsymbol{t}) = \{\boldsymbol{t}^* \in \mathcal{S}: G(\boldsymbol{t}^*) \geq G(\boldsymbol{t}))\}$ be the subsample space where the statistic would be equal or more extreme than the observed statistic given $\boldsymbol{t}$. Here ``more extreme'' suggests observations that are less likely under the null hypothesis. A natural choice of $G(\cdot)$ for these hypotheses is 
$G(\boldsymbol{t}) = \tau( \hat{\boldsymbol{\theta}}(\boldsymbol{t}) )$, where $\hat{\boldsymbol{\theta}}(\boldsymbol{t})$ denotes the vector of sample proportions, i.e., the maximum likelihood estimate of $\boldsymbol{\theta}$ given we observe $\boldsymbol{t}$. Then a p-value for testing $H_0$ can be defined as 
\begin{align}
p(\boldsymbol{t}, \psi_0) = \sup\left\{ \sum_{\boldsymbol{t}^* \in \overline{\mathcal{S}}(\boldsymbol{t})} f(\boldsymbol{t}^*, \boldsymbol{\theta}): \boldsymbol{\theta} \in \Theta_0(\psi_0)\right\}. \label{pvalue}
\end{align}

\begin{theorem}
The p-value defined in Equation \eqref{pvalue} is valid in the sense that 
\begin{eqnarray}
\sup\left\{\Pr[ p(\boldsymbol{T},\psi_0) \leq \alpha | \boldsymbol{\theta}]: \boldsymbol{\theta} \in \Theta_0(\psi_0)\right\} \leq \alpha \mbox{    for all $\alpha \in (0,1)$}.
\label{eq:valid.p.value}
\end{eqnarray}
\end{theorem}
All proofs are in the Appendix. 

Analogous theorems for the other direction one-sided hypotheses and the two-sided hypotheses are straightforward. Here is the notation for all these hypotheses.
Consider the class of hypotheses indexed by $\psi_0$ that was previously discussed, $H_{0 \ell}: \psi \leq \psi_0$ versus $H_{1 \ell}: \psi > \psi_0$, where now we add an index $\ell$ to denote ``lower'', foreshadowing the next section where we invert the associated p-value function to give a lower confidence limit. Let $p_{\ell}(\boldsymbol{t}, \psi_0)$
be the p-value associated with those hypotheses. Similarly, consider the class of hypotheses that switch direction of the inequalities, $H_{0 u}: \psi \geq \psi_0$ versus $H_{1 u}: \psi < \psi_0$. 
Its associated p-value, denoted $p_u(\boldsymbol{t},\psi_0)$, is obtained by switching the directions of the inequalities in the definitions of the sets $\Theta_0(\psi_0)$ and $\overline{\mathcal{S}}(\boldsymbol{t})$. A two-sided central p-value can be defined as  $p_c(\boldsymbol{t},\psi_0) =  \min\{1, 2 p_{\ell}(\boldsymbol{t},\psi_0), 2 p_u(\boldsymbol{t},\psi_0)\}$ and is a valid test of the hypothesis $H_{0c}: \psi = \psi_0$ versus $H_{1c}: \psi \neq \psi_0$.

\subsection{Confidence intervals}

Given the valid p-values $p_\ell(\boldsymbol{t}, \psi_0)$ and $p_u(\boldsymbol{t}, \psi_0)$, we can construct a valid $100 (1 - \alpha)\%$ confidence interval by inverting the p-value function. The lower limit associated with the $100(1-\alpha)\%$ one-sided confidence interval is
$\psi_{\ell[\alpha]}(\boldsymbol{t}) = \inf\{\psi: p_\ell(\boldsymbol{t}, \psi) > \alpha\}$. Similarly, let 
$\psi_{u[\alpha]}(\boldsymbol{t}) = \sup\{\psi: p_u(\boldsymbol{t},\psi) > \alpha\}$ be the upper limit associated with the other one-sided confidence interval. A $100(1 - \alpha)\%$ central confidence interval for $\psi$ is 
\begin{eqnarray}
    C(\boldsymbol{t}, 1 - \alpha) = \{\psi: \psi \geq \psi_{\ell[\alpha/2]}(\boldsymbol{t}) \mbox{ and } \psi \leq \psi_{u[\alpha/2]}(\boldsymbol{t})\}. \label{confint}
\end{eqnarray}

\begin{theorem}
    The confidence interval $C(\boldsymbol{t}, 1 - \alpha)$ in equation \eqref{confint} is valid, i.e., $\Pr\{\psi \in C(\boldsymbol{T}, 1-\alpha)\} \geq 1 - \alpha.$ It is also a central interval, i.e., the error is bounded at $\alpha/2$ on either side.
\end{theorem}

Since this confidence interval is an inversion of the two-sided test described in the previous section, we have inferential agreement between the two (c.f. Theorem 9.2.2 of \citet{casella2002statistical}).

\section{Implementation}
While the p-value is theoretically valid, it is difficult to compute it analytically in general. In the following, we propose a Monte Carlo estimation procedure for the exact p-value that we prove is consistent as the number of iterations increases. The confidence intervals are then based on the estimated p-values. We then describe the software implementation and how it can be used. 

\subsection{Monte Carlo implementation of exact p-values} \label{sec:monte}
It is difficult to compute $p(\boldsymbol{t}, \psi_0)$ analytically in general, so we propose the Monte Carlo estimation procedure in Algorithm \ref{palgo}.


\begin{algorithm}
\textbf{Input: } subsample space $\overline{\mathcal{S}}(\boldsymbol{t}) $, observed multinomial counts $\boldsymbol{t}$, null hypothesis value $\psi_0$, a distribution $H(\boldsymbol{t})$ on the $k$ probability simplices that may depend on $\boldsymbol{t}$, and a large integer $B$. \\
\textbf{Output: }$\hat{p}_B(\boldsymbol{t}, \psi_0)$ an estimated p-value for the null hypothesis parameter space $\Theta_0(\psi_0)$ \\
$\hat{p}_0(\boldsymbol{t}, \psi_0) \gets 0$; \\
\For{$i = 1, \ldots, B$}{
$\boldsymbol{\theta}^{(i)} \gets $ a random draw from $H(\boldsymbol{t})$; \\
\eIf{$\boldsymbol{\theta}^{(i)} \in \Theta_0(\psi_0)$}
{

$\hat{p}_i(\boldsymbol{t}, \psi_0) \gets \mbox{max}\left\{\hat{p}_{i-1}(\boldsymbol{t}, \psi_0), \sum_{\boldsymbol{t}^* \in \overline{\mathcal{S}}(\boldsymbol{t})} f(\boldsymbol{t}^*, \boldsymbol{\theta}^{(i)})\right\}$; 
}{
$\hat{p}_i(\boldsymbol{t}, \psi_0) \gets \hat{p}_{i-1}(\boldsymbol{t}, \psi_0)$; 
}
}

\If{$\hat{p}_B(\boldsymbol{t}, \psi_0) = 0$}
{
\mbox{ set $\hat{p}_B(\boldsymbol{t}, \psi_0)$ to NA;}
}
\textbf{Return}: $\hat{p}_B(\boldsymbol{t}, \psi_0)$;
\caption{Algorithm for computation of the p-value for the null $H_0: \boldsymbol{\theta} \in \Theta_0(\psi_0)$.  \label{palgo}}
\end{algorithm}

A default choice for the distribution $H(\boldsymbol{t})$ is the uniform distribution on the $k$ simplices, which does not depend on the data $\boldsymbol{t}$. Other choices are possible, as long as their support covers $\Theta_0(\psi_0)$, and $B * \Pr \left\{\boldsymbol{\theta}^{(i)} \in \Theta_0(\psi_0) \right\}$ is sufficiently large. If the estimated p-value after the for loop is $0$, this means that none of the $\boldsymbol{\theta}^{(i)} \in \Theta_0(\psi_0)$, no p-value is returned, and $B$ and/or $H(\boldsymbol{t})$ should be rethought. It is typically only of interest to accurately estimate $p(\boldsymbol{t}, \psi_0)$ when it is small, hence if at some point a $\hat{p}_i(\boldsymbol{t}, \psi_0)$ is found that exceeds some threshold, the loop can terminate early. Otherwise the resulting estimated p-value is $\hat{p}_B(\boldsymbol{t}, \psi_0)$.  

The following Lemma supports our main result of this section which is that the Monte Carlo implementation in Algorithm \ref{palgo} gives us a consistent estimate of the exact p-value in Equation \eqref{pvalue}.

\begin{lemma} \label{lemma1}
    Let $U_1, \ldots, U_m$ be independent identically distributed random variables from a space $\mathcal{U}$. Let $g: \mathcal{U} \rightarrow \mathbb{R}$ be continuous. Then
    \[
    \max\{g(U_1), \ldots, g(U_m)\} \rightarrow_p \sup_{u \in \mathcal{U}} g(u) \mbox{ as } m \rightarrow \infty.
    \]
\end{lemma}

\begin{theorem}
If $\Theta_0(\psi_0) \subseteq$ the support of $H(\boldsymbol{t})$
and $\Pr \left\{\boldsymbol{\theta}^{(i)} \in \Theta_0(\psi_0) \right\} > 0$, then 
\[
    \hat{p}_B(\boldsymbol{t}, \psi_0) \rightarrow_p p(\boldsymbol{t}, \psi_0)\mbox{ as }B \rightarrow \infty.
\]
\label{theorem:approx}
\end{theorem}

The condition for Theorem \ref{theorem:approx} to hold depends on specification of $H(\boldsymbol{t})$, the distribution that candidate $\boldsymbol{\theta}$ parameters are sampled from. If $\Theta_0(\psi_0)$ is a measurable subset of $\Theta$, i.e., it has nonzero volume, then any distribution that covers the full support of $\Theta$ will ensure both conditions are satisfied. 
There may be situations where $\Theta_0(\psi_0)$ is nonempty and also not measurable. That is why we need the second condition that $\Pr \left\{\boldsymbol{\theta}^{(i)} \in \Theta_0(\psi_0) \right\} > 0$. Consider the following example. Let $k = 1$, suppose $\tau(\theta) = \max_{\{i \in 1, \ldots, d\}}(\theta_{1i})$, and consider the null hypothesis $H_0: \psi \leq 1 / d$. The null parameter value is on the boundary of $\Psi$ and the null space $\Theta_0$ includes a single point: $\theta_0 =$ the $d$-length vector $(1/d, \ldots, 1/d)$. Our approximation algorithm will fail in this situation when using an $H(\boldsymbol{t})$ that is continuous and covers the full support because obtaining $\theta_0$ as a sample from a distribution with support on the probability simplex occurs with probability 0. However, we can compute the exact p-value here if we use an $H(\boldsymbol{t})$ that is discrete with support covering 
$\Theta_0$, because $\Theta_0$ is finite and therefore the supremum over that set is the maximum. In our implementation we allow for this possibility, and warn the users if they test a null hypothesis on the boundary of the $\Psi$ space. A detailed worked example of this is provided in the Supplementary Materials. 

Another example is $k=2$, $d_1=d_2=d$, and $\tau(\boldsymbol{\theta}) = \max_{i \in {1,\ldots, d}} | \theta_{1i} - \theta_{2i} |$,
and tests $H_0: \tau(\boldsymbol{\theta}) = \psi_0$.  In this case, $\Theta_0(\psi_0)$ is not finite, but its volume is $0$, so the default  $H(\boldsymbol{t})$ using the uniform simplex will have zero probability of sampling from the null hypothesis parameter space. In the Supplementary Materials, we show a distribution $H(\boldsymbol{t})$ that has a probability $1$ of sampling from $\Theta_0(\psi_0)$ when $\psi_0=0$, giving a tractable estimator of the p-value for $H_0: \tau(\boldsymbol{\theta})=0$ (equivalent to $H_0: \boldsymbol{\theta}_1=\boldsymbol{\theta}_2$);  however a distribution for sampling from   $\Theta_0(\psi_0)$ for general $\psi_0$ is not straightforward.

When the parameter space is unbounded, it may be advantageous to perform a transformation to a closed interval, e.g., inverse logit, before calculating confidence intervals. In practice, our suggested root-finding algorithm (for the confidence interval calculation) requires upper and lower bounds for the root search. Theoretically, with infinite Monte Carlo sampling there is only a single root (or a single contiguous set of roots) for either the upper or lower bound. This is because the size of $\Theta_0(\psi_0)$ grows as $\psi_0$ increases for $H_{0\ell}$ (and it grows as $\psi_0$ decreases for $H_{0u}$), hence by the definition of the one-sided p-values
(see Equation~\eqref{pvalue}), $p_{\ell}(\boldsymbol{t}; \psi_0)$ and $p_{u}(\boldsymbol{t}; \psi_0)$ are each monotonic in $\psi_0$. In practice, there may be multiple roots because of Monte-Carlo error. 

\subsection{Software implementation} \label{implementation}

Our proposed method is implemented in an \texttt{R} package \citep{R} called \texttt{xactonomial}, which is available on CRAN and at \url{https://sachsmc.github.io/xactonomial}. We have implemented Algorithm \ref{palgo} as described above as well as some strategies for speeding up the convergence of the p-value computation. For improvements in speed and memory safety, several functions have been implemented in Rust \citep{matsakis2014rust} and incorporated it into our R package using the framework described in \citet{cargo}. 

A prerequisite for Algorithm \ref{palgo} is an enumeration of the subsample space $\overline{\mathcal{S}}$, which can be computationally expensive. As a first step, we enumerate the sample space for each $T_j$, say $\mathcal{S}_j,$ for  $j \in 1, \ldots, k$. Then to obtain $\mathcal{S}$ we take all possible combinations of elements, one from each of the $\mathcal{S}_1, \ldots, \mathcal{S}_k$. The number of possible outcomes is (\citet{feller1968introduction}, page 38) 
\[
    \prod_{j = 1}^k \binom{n_j + d_j - 1}{d_j - 1},
 \]
which grows quite quickly in the number of samples, trials per sample, and dimension per sample. Enumerating those requires creation of a vector with $(\sum_j d_j) * \prod_{j = 1}^k \binom{n_j + d_j - 1}{d_j - 1}$ elements. In R, the maximum possible vector length is $2^{31}-1$, and of course it takes time to enumerate the elements and construct the matrix. Thus, there is a practical upper limit to the size of the problem, which to provide a point of reference, is nearly reached with about 13 trials in each of 3 samples each with dimension 4. We implemented the enumeration of the sample space in Rust, which dramatically speeds up the computation. 

The remaining computation of the p-value is a straightforward implementation of Algorithm \ref{palgo}. We implemented the random sampling from the simplex of dimension $d$ and calculation of the sum of multinomial probabilities in Rust. Although this approximate p-value converges to the exact p-value as the number of iterations grows, in some cases it may do so quite slowly, so we implemented two strategies for speeding up the convergence. 

The first strategy is to use a data-informed distribution $H(\boldsymbol{t})$. At each iteration we can sample $\theta_j$ independently from the Dirichlet distribution with shape parameter vector $1 + t_j$, where $t_j$ is the vector of observed counts for sample $j$. This is the posterior distribution for the multinomial parameters if we assume a uniform prior (Dirichlet prior with parameters $1$). This speeds up convergence because it concentrates the mass of the distribution around the observed data, while still covering the full support of the parameter space. 

The second strategy is to do gradient ascent. Let $\nabla \mathfrak{P}(\boldsymbol{\theta})$ denote the gradient of the probability $\sum_{t^* \in \overline{\mathcal{S}}} f(t^*, \boldsymbol{\theta})$ with respect to $\theta$. Then our p-value approximation algorithm incorporating gradient ascent is detailed in Algorithm \ref{algoascent}. It is unlikely that the p-value function is convex, so to protect against finding a local maximum we periodically refresh with a new draw from the base distribution $H(\boldsymbol{t})$.

\begin{algorithm}[ht]
\textbf{Input: } subsample space $\overline{\mathcal{S}}(\boldsymbol{t})$, observed multinomial counts $\boldsymbol{t}$, null hypothesis value $\psi_0$, integers $B, C_1, C_2$, an initial distribution $H(\boldsymbol{t})$ on the $k$ probability simplices, concentration rate $\gamma > 0$, learning rate $\lambda > 0$. \\
\textbf{Output: } $\hat{p}_B(\boldsymbol{t}, \psi_0)$ an estimated p-value for the null parameter space $\Theta_0(\psi_0)$ \\
$\boldsymbol{\hat{p}}\gets 0$; \\
$b \gets 1$; \\
1. \For{$i = 1, \ldots, C_1$}{
$\boldsymbol{\theta}^{(i)} \gets $ a random draw from $H(\boldsymbol{t})$; \\
\If{$\boldsymbol{\theta}^{(i)} \in \Theta_0( \psi_0)$}
{
\If{$\sum_{t^* \in \overline{\mathcal{S}}(\boldsymbol{t})} f(t^*, \boldsymbol{\theta}^{(i)}) > \max \boldsymbol{\hat{p}}$}{
$\boldsymbol{\hat{p}} \gets \mbox{append}(\boldsymbol{\hat{p}},  \sum_{t^* \in \overline{\mathcal{S}}(\boldsymbol{t})} f(t^*, \boldsymbol{\theta}^{(i)})$); \\
$\boldsymbol{\theta}^m = \boldsymbol{\theta}^{(i)}$; \\
}
}} 
\For{$j = 1, \ldots, C_2$}{
$b \gets b + 1$; \\
\If{$b \geq B$}{\eIf{$\max(\boldsymbol{\hat{p}}) = 0$}{\textbf{Return}: $\hat{p}_B(\boldsymbol{t}, \psi_0) =$ NA;}{
\textbf{Return}: $\hat{p}_B(\boldsymbol{t}, \psi_0) = \max(\boldsymbol{\hat{p}})$;}}
$\boldsymbol{\theta}^{(j)} \gets $ a random draw from the $k$ independent Dirichlet distributions with concatenated vector of shape parameters
\[
    \frac{\gamma * (\theta^m + \lambda * \nabla \mathfrak{P}(\theta^m))}{ \boldsymbol{1}^\top(\theta^m + \lambda * \nabla \mathfrak{P}(\theta^m))}
\] 
where $\boldsymbol{1}$ is the vector of 1s the same length as $\theta^m$. \\
\If{$\boldsymbol{\theta}^{(j)} \in \Theta_0(\psi_0)$}
{
\If{$\sum_{t^* \in \overline{\mathcal{S}}(\boldsymbol{t})} f(t^*, \boldsymbol{\theta}^{(j)}) > \max \boldsymbol{\hat{p}}$}{
$\boldsymbol{\hat{p}} \gets \mbox{append}(\boldsymbol{\hat{p}},  \sum_{t^* \in \overline{\mathcal{S}}(\boldsymbol{t})} f(t^*, \boldsymbol{\theta}^{(j)})$); \\
$\boldsymbol{\theta}^m \gets \boldsymbol{\theta}^{(j)};$ \\
}
}
}
\textbf{Go to: } 1.
\caption{Algorithm to approximate p-values that incorporates gradient ascent.
\label{algoascent}}
\end{algorithm}

\begin{figure}[ht]
\centering
\includegraphics[scale = .8]{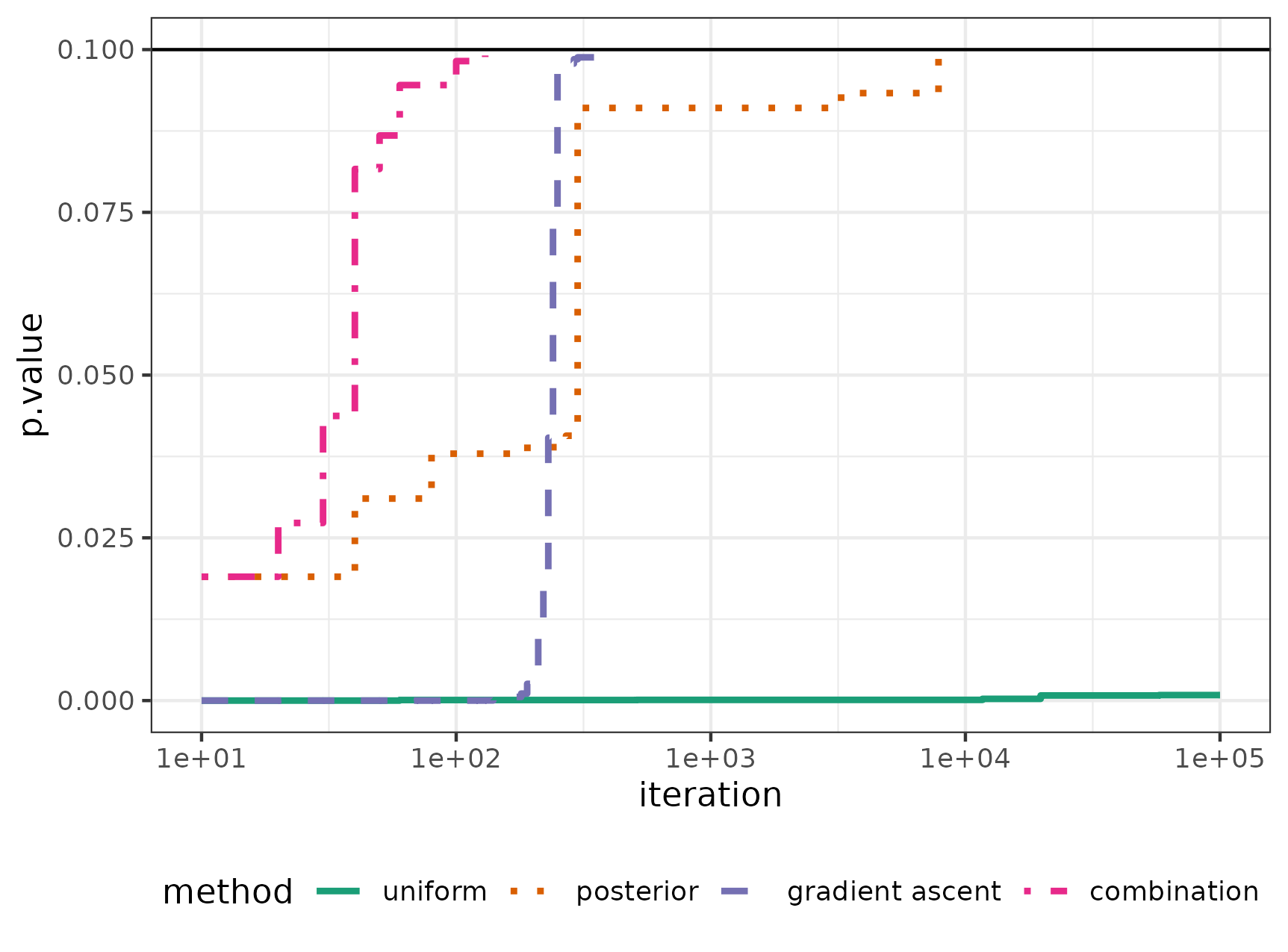}
\caption{\label{ascent} P-values as the iterations grow for uniform = sampling from the uniform simplex, posterior = sampling from the posterior Dirichlet assuming a uniform prior, gradient ascent = using gradient ascent, and combination = using a combination of gradient ascent and posterior Dirichlet sampling. The true p-value is 0.1. }
\end{figure}

To illustrate how these modifications can dramatically improve the speed at which the procedure finds a larger p-value, consider the following example, which was helpfully pointed out by a reviewer. Suppose $k = 1$, $d = 10$, $n = 20$, $\tau(\boldsymbol{\theta}) = {\theta}_1$, and we want to test $H_0: \theta_1 \leq 0.1^{(1/20)}$. If we observe a sample where all 20 trials are in category 1, then the true p-value is 0.1. Algorithm \ref{palgo} with uniform sampling is slow to converge in this setting, but gradient ascent and sampling from a non-uniform Dirichlet distribution informed by the data dramatically improves performance. This is illustrated in Figure \ref{ascent}, which shows the approximate p-value as the iterations increase for the different algorithms. 

Confidence intervals are computed by finding the values of $\psi_0$ that solve $p_{\ell}(\boldsymbol{t}, \psi_0) - \alpha/2 = 0$ and $p_u(\boldsymbol{t}, \psi_0) - \alpha/2 = 0$, or using a boundary of the $\psi$ parameter space if no solution exists. To minimize the number of p-value function evaluations during this process, we use the interpolate, truncate, project (ITP) root finding algorithm \citep{oliveira2020enhancement}, which converges with fewer function evaluations compared to the Brent method that is implemented in the R function \texttt{uniroot}. We implemented the ITP algorithm in base R, though an add on package called \texttt{itp} also exists \citep{itp}. To further reduce computation time, we have an optional argument that terminates the inner loop of the p-value computation early if the current p-value is larger than the given parameter.

\subsection{Description of the software package} \label{package}

The R package's primary feature is its eponymous function. Users must provide three required arguments to the \texttt{xactonomial} function: the data which must be provided as a list of $k$ vectors each of which represents the vector of observed counts in each cell for each of the $k$ samples, the function $\tau(\boldsymbol{\theta})$ that maps the vector of multinomial parameters to the real line, and a vector giving the lower and upper limits of the possible range of $\psi$. There are a number of optional arguments, including the null value to test, the direction of the test, whether to compute a confidence interval, the confidence level, and parameters to control the algorithms. The maximum number of iterations for the p-value computation ($B$) is determined by two arguments, the maximum number of iterations, and the chunk size, that is, the size of the sample of random draws for $\boldsymbol{\theta}$ at each iteration. 

It only requires a few lines of \texttt{R} code to run: 

\begin{verbatim}
tau_bc <- function(theta) {
  
  theta1 <- theta[1:4]
  theta2 <- theta[5:8]
  sum(sqrt(theta1 * theta2))
  
}

set.seed(2024)
data <- list(T1 = c(6,1,2,1), T2 = c(1,1,5,3))
xactonomial::xactonomial(data, f_param = tau, psi0 = 0.75, psi_limits = c(0,1))

## 	Monte-Carlo multinomial test
## 
## data:  data
## p-value = 0.2662
## alternative hypothesis: true psi0 is not equal to 0.75
## 95 percent confidence interval:
##  0.6325000 0.9971308
## sample estimates:
## [1] 0.8343818
\end{verbatim}

When the null hypothesis is on the boundary of the $\Psi$ space (i.e., all $\boldsymbol{\theta} \in \Theta_0(\psi_0)$ have $\tau(\boldsymbol{\theta})=\psi_0$, and either $\psi_0 =\min \Psi$ or $\psi_0 = \max \Psi$), users will receive a warning that, if possible, they should provide the optional argument \texttt{theta\_null\_points} which is the matrix containing the finite set of $\boldsymbol{\theta}$ values at which $\tau(\boldsymbol{\theta})$ equals the boundary. If the set is finite and provided then the p-value will be truly exact, as described at the end of Section \ref{sec:monte}. If no points are sampled from the null space, then users will be warned and receive a p-value of NA. This issue can occur if the null space is infinite but has volume 0, or if there are not enough iterations. The latter problem can be solved by specifying a function to the argument \texttt{theta\_sampler} to sample direcly from the null space, if possible.

The \texttt{xactonomial} function returns an object of class \texttt{htest}, to be consistent with other hypothesis testing functions in R. It is a list that contains relevant information about the test and confidence interval which is printed nicely and works with other functions that operate on \texttt{htest} objects. It also includes a list element that contains the sequence of p-values obtained at each iteration. 

\begin{figure}[ht]
\centering
\includegraphics[scale=.6]{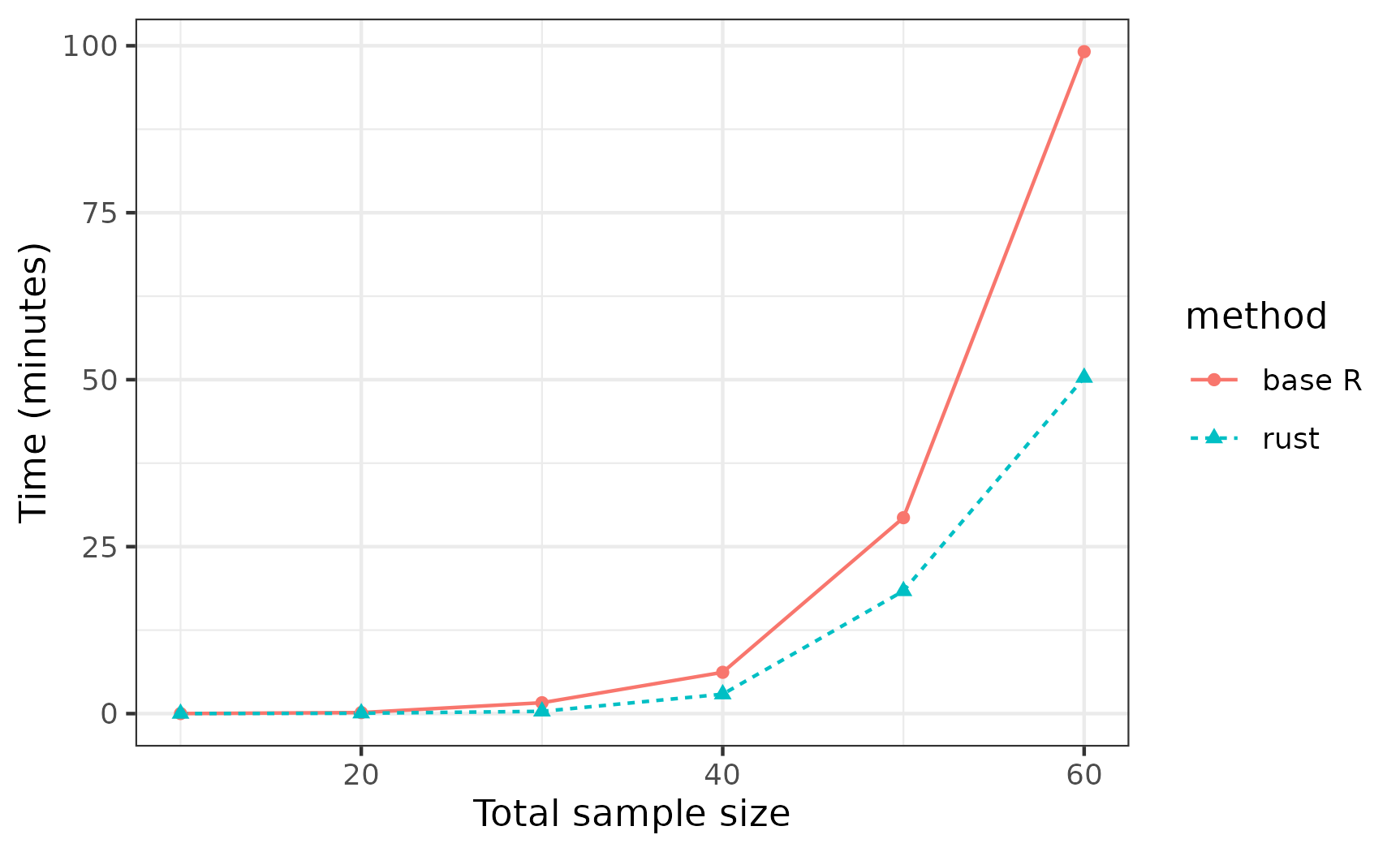}
\caption{Comparison of computation speed of p-values between the base R and Rust implementations. The y-axis is the time to compute the sample space, and run 100 iterations of the p-value calculation taking 100 uniform samples from $\Theta$ at each iteration in the Bhattacharyya coefficient setting with $k = 2$, $d = 4$, and equal sample sizes in the two groups. \label{speedtest}}
\end{figure}

As mentioned above, some of the underlying functions are implemented in \texttt{Rust}, and incorporated into our package using the framework described by \citet{cargo}. This led to dramatic gains in speed and memory usage, allowing us to do the computations in a reasonable amount of time for total sample sizes up to about 60. This is illustrated in Figure \ref{speedtest}, which shows the timings of the p-value computation after having enumerated the subsample space with the base R implementation and the rust implementation. These experiments were done with the Bhattacharyya coefficient setting with $k = 2$, $d = 4$, and equal sample sizes in the two groups. The calculations were done on a laptop running Ubuntu 22 with an AMD Ryzen 7 and 32 GB of RAM. 

\section{Simulations} \label{simulations}
We examine the numeric performance of our method in comparison to the nonparametric bootstrap. We do this using two examples of functions, the Bhattacharyya coefficient, and the maximum of multinomial parameters. In all cases we used 500 bootstrap replicates, 2000 simulation replicates, and used a maximum of 10000 iterations of our method. Code and instructions to reproduce the numerical experiments is available at \url{https://sachsmc.github.io/xactonomial-paper}. The simulations were run on a computing cluster with Slurm and took approximately 7200 cpu-hours to complete.

\subsection{Bhattacharyya coefficient}

Suppose $d_j = d$ and $n_j = n$ for all $j = 1, \ldots, k$, and that we observe the multinomial random vectors $T_1, \ldots, T_k$ which have parameters $\boldsymbol{\theta}_1, \ldots, \boldsymbol{\theta}_k$. Let 
\[
\tau(\boldsymbol{\theta}) = \sum_{i = 1}^d \left({\prod_{j = 1}^k\theta_{ji}}\right)^{1/k},
\]
which is the extension of the Bhattacharyya coefficient to more than 2 distributions, as proposed in \citet{kang2015n}. We examine the performance of our proposed method in comparison to the nonparametric bootstrap in terms of empirical 95\% confidence interval coverage, for different values of $d, n$ and $\boldsymbol{\theta}$. The results are shown in Table \ref{bhatty}. The nonparametric bootstrap, which is an asymptotic method, unsurprisingly has very poor coverage at small to moderate sample sizes, while our proposed method appears to have at least nominal coverage (within simulation error) in all cases (although in many cases the coverage is conservative). We also report the type I error rate and average width of the confidence intervals for our method. The simulated type I error rate is always below or close to the nominal level of $5\%$. The average width is quite wide for small sample sizes, but for sample sizes of 10 or 20, the confidence intervals can be quite informative, with average width as low as 20\% of the parameter range (which is 0 to 1) or narrower. 

\begin{table}[ht]
\caption{Bhattacharyya coefficient simulation: Empirical coverage (percent) of nominal 95\% confidence intervals  using the proposed method in comparison to the nonparametric bootstrap for the Bhattacharyya coefficient under different sample sizes, dimensions, and true values. n = sample size, k = samples, d = dimension. Type I error as the proportion of two-sided tests rejected for the null at the true parameter value, and average width of the confidence intervals are presented for our method.
\label{bhatty}}
\centering
\begin{tabular}{rr|r|rrr}
  \hline
  & & \multicolumn{2}{|c}{coverage (\%)} & type I error & avg. width \\
n / k / d &  true value &  bootstrap & \multicolumn{3}{|c|}{xactonomial} \\ 
  \hline
   5 / 4 / 3 & 0.85 & 8.9  & 99.2 & 0.016 & 0.69 \\ 
   5 / 5 / 3 & 0.97 & 0.1  & 95.0 & 0.014 & 0.52 \\ 
  10 / 2 / 3 & 0.73 & 64.8 & 97.5 & 0.048 & 0.42  \\ 
  10 / 2 / 3 & 0.77 & 55.4 & 96.4 & 0.018 & 0.53  \\ 
  10 / 2 / 4 & 0.98 & 43.9 & 96.5 & 0.028 & 0.29  \\ 
  10 / 2 / 4 & 0.99 & 67.0 & 97.9 & 0.044 & 0.25  \\ 
  10 / 3 / 3 & 0.81 & 40.5 & 97.9 & 0.028 & 0.50  \\ 
  20 / 2 / 3 & 0.73 & 76.9 & 95.2 & 0.028 & 0.55  \\ 
  20 / 2 / 3 & 0.77 & 72.9 & 94.8 & 0.056 & 0.40  \\ 
  20 / 2 / 4 & 0.98 & 71.2 & 98.4 & 0.012 & 0.16  \\ 
  20 / 2 / 4 & 0.99 & 83.8 & 99.1 & 0.008 & 0.12  \\ 
   \hline
\end{tabular}
\end{table}

\subsection{Maximum of parameters}

Suppose we observe a single multinomial vector with $n$ trials, dimension $d$, and parameter vector $\boldsymbol{\theta}$. Let $\tau(\boldsymbol{\theta}) = \max_{\{i \in 1\ldots, d\}}\{\theta_{1i}\}$. We consider different values for $n$, $d$, and the vector of parameters $\boldsymbol{\theta}$, over 2000 simulation replicates and present the empirical confidence interval coverage of our method in comparison to the nonparametric bootstrap. In situations where two or more of the elements of $\boldsymbol{\theta}$ are equal to each other, we expect the nonparametric bootstrap to perform poorly, even in large sample sizes, because of the lack of differentiability. The results are shown in Table \ref{lbsims}. The nonparametric bootstrap has less than nominal coverage at small to moderate sample sizes, whereas our method has nominal to slightly conservative coverage in these settings. Again the type I error is less than nominal of 5\%, even in the difficult settings where $\psi_0$ is on the boundary of the parameter space. The average width shows that they can be quite informative. 

\begin{table}[ht]
\caption{Maximum Simulation: Empirical coverage (percent) of the nominal 95\% confidence intervals of the proposed method in comparison to the nonparametric bootstrap for the maximum multinomial parameter under different sample sizes and true values. Type I error as the proportion of two-sided tests rejected for the null at the true parameter value, and average width of the confidence intervals are presented for our method.\label{lbsims}}
\centering
\begin{tabular}{rr|r|rrr}
  \hline
  &&\multicolumn{2}{|c}{coverage (\%)} & type I error (\%) & avg. width \\
Sample size (n) & $\theta$   & bootstrap & \multicolumn{3}{|c|}{xactonomial} \\ 
  \hline
10 & $(0.2, 0.5, 0.3)$         & 96.7& 98.7 &  0.011 & 0.49 \\
20 & $(0.4, 0.4, 0.2)$         & 85.5& 97.5 &  0.013 & 0.37 \\
30 & $(0.3, 0.3, 0.25, 0.15)$  & 68.5& 99.0 &  0.013 & 0.30 \\
40 & $(0.3, 0.3, 0.25, 0.15)$  & 75.5& 98.5 &  0.012 & 0.27 \\
10 & $(1/3, 1/3, 1/3)$         & 0.0 & 95.5 &  0.010 & 0.46 \\
20 & $(1/3, 1/3, 1/3)$         & 0.0 & 96.7 &  0.016 & 0.33 \\
30 & $(1/3, 1/3, 1/3)$         & 8.2 & 95.4 &  0.021 & 0.27 \\
40 & $(1/3, 1/3, 1/3)$         & 0.0 & 96.8 &  0.015 & 0.23 \\
   \hline
\end{tabular}
\end{table}

\section{Real data example} \label{example}

\citet{du2015randomized} report the results of a randomized trial to evaluate the effect of peanut consumption during infancy on the development of peanut allergy. The publicly available trial data were downloaded from the Immune Tolerance Network TrialShare website on 2020-06-15 (\url{https://www.itntrialshare.org/}, study identifier: ITN032AD).
The trial had a two-armed randomization with one arm assigned to consume a certain amount of peanut powder per week and one arm assigned to avoid consuming peanut powder. In this study, 640 participants between 4 months and 11 months of age were randomized to either consume peanuts or avoid peanuts until the age of 60 months. The randomization was stratified on positivity of a baseline skin prick test for peanut protein. Those in the consumption arm were instructed to consume at least 6 grams per week. At the end of the study, the outcome allergy to peanuts was assessed using a oral food challenge. Compliance with the assigned intervention was assessed weekly by using a food frequency questionnaire, and by manual inspection of the infants' cribs for peanut crumbs in a randomly selected subset of participants.

It is of interest to estimate the causal risk difference of peanut exposure on the risk of peanut allergy in this study. This is a classic example of a randomized trial with noncompliance, as some of the participants assigned to consume peanuts actually avoided them. We are interested in the parameter $\beta = \Pr\{Y(X = 1) = 1\} - \Pr\{Y(X = 0) = 1\}$, where the notation $Y(X = x)$ denotes potential outcomes which represent the random variable of peanut allergy $Y$ if everyone in the population were intervened upon to have been exposed to peanuts $(X = 1)$ or avoided peanuts $(X = 0)$ \citep{rubin1974estimating}. Without any other assumptions, $\beta$ is not identified, however tight bounds have been derived for this setting in terms of observable conditional probabilities of the form $p_{xy\cdot z} = \Pr\{X = x, Y = y | Z = z\}$, where $Z$ indicates assigned treatment \citep{balke1995thesis, Balke97, swanson2018partial}. Let the bounds be $\psi_{\ell b} \leq \beta \leq \psi_{ub}$, where the lower bound is 
\begin{eqnarray}
 \beta &\geq& \mbox{max} \left. \begin{cases}   
   -1 + p_{00\cdot 1} + p_{11\cdot 1},\\ 
    -1 + p_{00\cdot 1} + p_{11\cdot 0}, \\
   -1 + p_{00\cdot 0} + p_{11\cdot 0},\\ 
 -1 + p_{00\cdot 0} + p_{11\cdot 1},\\ 
   -2 + 2p_{00\cdot 0} + p_{01\cdot 1} + p_{11\cdot 0} + p_{11\cdot 1},\\ 
   -2 + p_{00\cdot 0} + p_{00\cdot 1} + p_{10\cdot 0} + 2p_{11\cdot 1},\\ 
     -2 + 2p_{00\cdot 1} + p_{01\cdot 0} + p_{11\cdot 0} + p_{11\cdot 1},\\ 
   -2 + p_{00\cdot 0} + p_{00\cdot 1} + p_{10\cdot 1} + 2p_{11\cdot 0}.
   \end{cases}  \label{lower1} \right\}  = \psi_{\ell b}
\end{eqnarray}
and the upper bound is
\begin{eqnarray*}
 \beta &\leq& \mbox{min} \left. \begin{cases}   1 - p_{10\cdot 1} - p_{01\cdot 0},\\ 
    1 - p_{10\cdot 1} - p_{01\cdot 1},\\ 
   1 - p_{10\cdot 0} - p_{01\cdot 0},\\ 
      1 - p_{10\cdot 0} - p_{01\cdot 1},\\ 
   2 - 2p_{10\cdot 1} - p_{01\cdot 0} - p_{01\cdot 1} - p_{11\cdot 0},\\ 
   2 - p_{00\cdot 1} - p_{10\cdot 0} - p_{10\cdot 1} - 2p_{01\cdot 0},\\ 
 2 - 2p_{10\cdot 0} - p_{01\cdot 0} - p_{01\cdot 1} - p_{11\cdot 1},\\ 
   2 - p_{00\cdot 0} - p_{10\cdot 0} - p_{10\cdot 1} - 2p_{01\cdot 1}. 
\end{cases}   \label{upper1} \right\}  = \psi_{ub}
\end{eqnarray*}
Let 
\[
T_0 = \left(
\begin{array}{c} 
\# \left\{ X = 0, Y = 0 | Z = 0 \right\} \\
\# \left\{ X = 1, Y = 0 | Z = 0 \right\} \\
\# \left\{ X = 0, Y = 1 | Z = 0 \right\} \\
\# \left\{ X = 1, Y = 1 | Z = 0 \right\}
\end{array}
\right)
\mbox{ and } 
T_1 = \left(
\begin{array}{c} 
\# \left\{ X = 0, Y = 0 | Z = 1 \right\} \\
\# \left\{ X = 1, Y = 0 | Z = 1 \right\} \\
\# \left\{ X = 0, Y = 1 | Z = 1 \right\} \\
\# \left\{ X = 1, Y = 1 | Z = 1 \right\} 
\end{array}
\right).
\]
$T_0$ and $T_1$ can be viewed as independent multinomial random variables with probabilities $(p_{00\cdot 0}, p_{10\cdot 0}, p_{01\cdot 0}, p_{11\cdot 0})$ and $(p_{00\cdot 1}, p_{10\cdot 1}, p_{01\cdot 1}, p_{11\cdot 1})$, respectively. The expressions for the lower and upper bounds are thus real-valued functions of these multinomial probabilities, and hence we can apply our method. 

In this example, we examine the subgroup of participants who were positive on a skin-prick test for peanut allergy at baseline, stratified by sex. The data are reported as cell counts and percents in Table \ref{peanutgramtable}, by sex and assigned treatment. The results of the analysis are in Table \ref{peanutresults}, which shows the estimated bounds, exact inference, bootstrap inference, and the estimated terms that make up the bounds expressions. The bootstrap inference is less conservative compared to our method, but as our simulations show, we often attain nominal coverage while the bootstrap potentially has under-coverage. The last 8 rows of the Table \ref{peanutresults} shows the estimated terms that go into the bounds expressions, with the lower bound being the maximum of the 8 terms and the upper bound the minimum of the 8 terms. For the girls stratum, there are 2 terms that equal the maximum for the lower bound and 2 terms that equal the minimum for the lower bound, while for boys, they are close but not equal. While this does not imply that the true probabilities are equal, it does raise concerns about the validity of the bootstrap confidence intervals. 

The resulting estimated change in risk of peanut allergy development is between $-$34 and $-$16\% in boys with a combined (95\% CI: $-$71 to 24\%) and between $-$24 and $-$18\% in girls with a combined (95\% CI: $-$50\% to 34\%). The causal null hypothesis is that there is no difference, which would be falsified by the bounds if the lower bound is significantly greater than 0, or if the upper bound is significantly less than 0. The P-values for testing the null hypotheses were not statistically significant at the 0.05 level in either boys or girls, which is not surprising given the combined 95\% CIs.

\begin{table}[ht]
\caption{Counts and percents (of each cell within arm) of treatment taken and outcomes, by assignment group and sex, in the high-risk subgroup in the peanut allergy study \citep{du2015randomized}. \label{peanutgramtable}}
\centering
\begin{tabular}{l|rr|rr}
\multicolumn{5}{c}{Boys} \\
\hline
& \multicolumn{2}{r|}{Avoidance Arm ($n = 32$)} & \multicolumn{2}{c}{Consumption Arm ($n = 28$)} \\
Outcome & Tolerant & Allergic &  Tolerant & Allergic \\
 \hline
 Treatment taken & & & & \\
 Peanut avoidance & 20 (62.5\%) & 12 (37.5\%)  & 2 (7.1\%) & 3 (10.7\%) \\
 Peanut consumption & 0 (0.0\%) & 0 (0.0\%)  & 22 (78.6\%) & 1 (3.6\%) 
 \end{tabular}
 
\begin{tabular}{l|rr|rr}
\multicolumn{5}{c}{Girls} \\
\hline
& \multicolumn{2}{r|}{Avoidance Arm ($n = 17$)} & \multicolumn{2}{c}{Consumption Arm ($n = 19$)} \\
Outcome & Tolerant & Allergic &  Tolerant & Allergic \\
 \hline
 Treatment taken & & &  & \\
 Peanut avoidance & 13 (76.5\%) & 4 (23.5\%)  & 0 (0.0\%) & 1 (5.3\%) \\
 Peanut consumption & 0 (0.0\%) & 0 (0.0\%)  & 18 (94.7\%) & 0 (0.0\%) \\
 \hline
 \end{tabular}
\end{table}

\begin{table}[ht]
\caption{\label{peanutresults} Results of the peanut trial analysis}
\centering
\begin{tabular}{|l|c|c|c|c|}
\hline
 & \multicolumn{2}{|c|}{Boys} & \multicolumn{2}{|c|}{Girls} \\
\hline
& lower bound & upper bound & lower bound & upper bound  \\
estimate & $-0.34$ & $-0.16$ & $-0.24$ & $-0.18$ \\
exact 95\% CI & $(-0.71, 0.05)$ & $(-0.50, 0.24)$& $(-0.71, 0.25)$ & $(-0.59, 0.34)$\\
null hypothesis & $\psi_{\ell b} \leq 0$ & $\psi_{ub} \geq 0$ & $\psi_{\ell b} \leq 0$ & $\psi_{ub} \geq 0$ \\
exact 1-sided p-value & 0.99 & 0.22 & 0.99 & 0.23 \\
bootstrap 95\% CI &  $(-0.53, -0.15)$ & $(-0.39, 0.07)$ &  $(-0.47, -0.01)$ & $(-0.41, -0.01)$\\
\hline 
& \multicolumn{4}{c}{Individual bounds terms} \\
& $-0.34$ & $-0.16$ & $-0.24$ & $-0.18$ \\
& $-0.61$ & $-0.05$ & $-0.42$ & $-0.18$ \\
& $-1.23$ & $ 0.39$ & $-1.24$ & $ 0.58$ \\
& $-0.89$ & $ 0.11$ & $-1.00$ & $ 0.00$ \\
& $-0.38$ & $ 0.62$ & $-0.24$ & $ 0.76$ \\
& $-0.52$ & $ 1.48$ & $-0.29$ & $ 1.71$ \\
& $-1.45$ & $ 0.38$ & $-1.76$ & $ 0.18$ \\
& $-0.93$ & $ 0.89$ & $-1.00$ & $ 0.95$ \\
\hline
\end{tabular}    
\end{table}

\section{Discussion} \label{discussion}
As we have demonstrated, particularly in small samples, inference for a real-valued, continuous function of probabilities in the $k$-sample multinomial problem can be difficult. Although we prove that our proposed theoretical method provides an exact p-value for testing and correct coverage for confidence intervals, we also demonstrate that it is infeasible unless the null parameter space is finite. Instead, we propose a computational approximation, which we prove to be consistent as the number of iterations of the approximation grows. Although the approximate p-value will always be less than the exact p-value, the approximation can be very good and substantially better than inference based on the bootstrap. We proposed and implemented some methods for speeding up the approximation. 

We demonstrated that this approximation has good finite sample properties via simulations in settings where bootstrap methods fail. We highlight a number of settings, such as measure zero, but non-empty, null spaces, where the approximation can be made exact with user intervention. This is notably a setting where all other methods would fail to give correct coverage or type one error rate. We provide the R package \texttt{xactonomial}, for easy use of the method, which is now available on CRAN. 

Our proposed method is general, allowing for any real-valued parameter of interest, only constrained by the nature of the parameter space. However, when the setting of interest constrains the parameter space or the sample space, it may be possible to use this information to make inference more efficient. An example is causal bounds, which are often derived under a set of counterfactual assumptions that may restrict the possible space of the multinomial parameters. Leveraging such constraints effectively in estimation and inference is a future area of research for the authors.


\section*{Supplementary Material}
The R package implementing the method is available on CRAN at \url{https://cran.r-project.org/package=xactonomial} and at \url{https://sachsmc.github.io/xactonomial}. Code relating to this paper is available at  \url{https://sachsmc.github.io/xactonomial-paper}. It includes code for the simulation study, real data example, and the ``difficult'' examples described in the main text. The examples are also included as a PDF with the output and detailed descriptions of the settings. 

\section*{Acknowledgements}
This work utilized the computational resources of the NIH HPC Biowulf cluster. (\url{https://hpc.nih.gov}), and thanks to Vivian Callier for help with interfacing with it. 

EEG was partially supported by a grant from Novo Nordisk fonden NNF22OC0076595. EEG and MCS were partially supported by the Pioneer Center for SMARTbiomed. 

\selectlanguage{english}
\bibliographystyle{unsrtnat}
\bibliography{references.bib%
}

\appendix

\section*{Appendix}

\begin{proof}{of Theorem 1.}
Suppose there are 
$M$ elements in $\mathcal{S}$, and those $M$ possible values of $\boldsymbol{T}$ create 
$m \leq M$ possible values of  $G(\boldsymbol{T})$, say $\gamma_1 > \gamma_2 > \cdots > \gamma_m$.
(We order the $\gamma_j$  from largest to smallest so that the p-values will be from smallest to largest.) 
Let $\boldsymbol{t}_j^*$ be a value of $\boldsymbol{T}$ such that $G(\boldsymbol{t}_j^*)=\gamma_j$.
Let 
 $\bar{\mathcal{S}}(\boldsymbol{t}_j^*) = \bar{\mathcal{S}}_j = \left\{ {\boldsymbol{t}} \in \mathcal{S} : G(\boldsymbol{t}) \geq G(\boldsymbol{t}^*_j) \right\}$.
Then we can partition the support of $\boldsymbol{T}$ into $m$ ordered sets, 
\begin{eqnarray*}
 \bar{\mathcal{S}}_1 \subset \bar{\mathcal{S}}_2 \subset \cdots \subset \bar{\mathcal{S}}_m \equiv \mathcal{S}
\end{eqnarray*} 
Let $\bar{\mathcal{S}}_0 = \emptyset$.
Then by the definition of
\begin{eqnarray*}
p_j = p(\boldsymbol{t}_j^*,\psi_0) & = & 
\sup \left\{\sum_{t^* \in \bar{\mathcal{S}}_j} f(\boldsymbol{t}^*,\theta): \boldsymbol{\theta} \in \Theta_0(\psi_0)\right\}
= \sup\left\{\Pr \left[ \boldsymbol{T} \in \bar{\mathcal{S}}_j \vert \boldsymbol{\theta} \right]: \boldsymbol{\theta} \in \Theta_0(\psi_0)\right\}
\end{eqnarray*}
 and there are $m$ possible values for the p-value, $0 < p_1 < p_2 < \cdots < p_m=1$. 
  
Without loss of generality, assume  $\alpha \in [p_{j-1},p_{j})$ with $p_0=0$, for some
arbitrary $j \in \left\{ 1,\ldots,m \right\}$, then 
\begin{eqnarray*}
\sup\left\{ \Pr[ p(\boldsymbol{T},\psi_0) \leq \alpha | \theta] : \theta \in \Theta_0(\psi_0) \right\} 
& = & \sup\left\{ \Pr[ p(\boldsymbol{T},\psi_0) \leq p_{j-1} | \theta] : \theta \in \Theta_0(\psi_0)\right\} \\
& = & \sup\left\{ \Pr[ \boldsymbol{T} \in  \bar{\mathcal{S}}_{j-1} | \theta]: \theta \in \Theta_0(\psi_0)\right\} =  p_{j-1} \leq \alpha.
\end{eqnarray*}
\end{proof}

\begin{proof}{of Theorem 2.}
\begin{eqnarray*}
    \Pr\{\psi \in C(\boldsymbol{T}, 1 - \alpha)\} = \\ 
    1 - \Pr\{\psi < \psi_{\ell[\alpha/2]}(\boldsymbol{T})\} - \Pr\{\psi > \psi_{u[\alpha/2]}(\boldsymbol{T})\}
    \geq 1 - \alpha.
\end{eqnarray*}
\end{proof}

%

\begin{proof}{of Lemma \ref{lemma1}.}
Let $F_g(x)$ denote the cumulative distribution function of $g(U_1)$. Then 
    \begin{align*}
    & \Pr\left\{\vert\max\{g(U_1), \ldots, g(U_m)\} - \sup_{u \in \mathcal{U}} g(u)\vert < \varepsilon\right\} = \Pr\left\{\max\{g(U_1), \ldots, g(U_m)\} > \sup_{u \in \mathcal{U}} g(u) - \varepsilon\right\} = \\
    & \left(1 - F_g(\sup_{u \in \mathcal{U}} g(u) - \varepsilon)\right)^m,
    \end{align*}
   from basic results on the distribution of the order statistics of a sample of size $m$ from the distribution of $g(U)$. Since $\varepsilon > 0$ we have $F_g(\sup_{u \in \mathcal{U}} g(u) - \varepsilon) > 0$ and hence the expression being raised to the power $m$ is less than 1. Thus the probability $\rightarrow 0 \mbox{ as } m \rightarrow \infty$.
\end{proof}

\begin{proof}{of Theorem~\ref{theorem:approx}}

Let the sample from $H$ be $\theta^{(1)},\ldots, \theta^{(B)}$, and let
 $M = \sum_{i=1}^{B} I \left\{ \theta^{(i)} \in \Theta_0(\psi_0) \right\}$ be the number of the samples in the null hypothesis space.
 $E(M) \rightarrow \infty$ as $B \rightarrow \infty$ because $\Pr[ \theta^{(i)} \in \Theta_0(\psi_0)]>0$.
Let $i_1,\ldots,i_M$ be the values of $i$ 
where $\theta^{(i)} \in \Theta_0(\psi_0)$.
Then Theorem \ref{theorem:approx} follows from Lemma \ref{lemma1} with
$U_j=\theta^{(i_j)}$, $m=M$, and 
\[
g(U) = \sum_{t^* \in \overline{\mathcal{S}}(\boldsymbol{t})} f(t^*, U).
\]
\end{proof}
\end{document}